\documentstyle[aps,eqsecnum,floats]{revtex}

\def\be{\begin{equation}}
\def\en{\end{equation}}
\def\bear{\begin{eqnarray}}
\def\enar{\end{eqnarray}}
\def\beas{\begin{eqnarray*}}
\def\enas{\end{eqnarray*}}

\def\dsp{\displaystyle}
\def\ptl{\partial}

\def\itwo{\dsp{i\over 2}}
\def\im{{\rm Im}}
\def\re{{\rm Re}}
\newcommand{\half}{\frac{1}{2}}
\def\ep{\epsilon}

\def\CA{{\cal A}}

\def\CC{{\cal C}}
\def\CD{{\cal D}}

\def\buildchar#1#2#3{\null \! \mathop {\vphantom {#1}
\smash #1}\limits ^{#2}_{#3}\!\null }
\def\ut#1{\buildchar{#1}{}{^\sim}\/}
\def\dtri{\tilde{E}}

\def\den{e}
\def\been{\begin{enumerate}}
\def\enen{\end{enumerate}}

\def\det{ \mbox{det} \,}
\def\rank{\mbox{rank} \,}
\newtheorem{Def}{Definition}
\begin{document}
\draft

\title{
Constructing hyperbolic systems 
in the Ashtekar formulation
of general relativity \footnote{This article was published as   
Int. J. Mod. Phys. D 9 (2000) 13.   
We corrected typo in eq.(\ref{symhypIIIa_eigen}) and added a note
in the end 
in this online version. (gr-qc/9901053)}
}

\author{Gen Yoneda \cite{Email-yone}}
\address{
Department of Mathematical Science, Waseda University,
Okubo 3-4-1, Shinjuku, Tokyo 169-8555, Japan}
\author{Hisa-aki Shinkai  \cite{Email-his}}
\address{
Center for Gravitational Physics and Geometry, 
104 Davey Lab., Department of Physics, \\
The Pennsylvania State University,
University Park, Pennsylvania 16802-6300, USA}
\date{November 5, 1999  (revised version)}
\maketitle
\begin{abstract}

Hyperbolic formulations of the equations of motion are essential 
technique for proving the well-posedness of the Cauchy problem of 
a system, and are also helpful for implementing stable long time 
evolution in numerical applications.
We, here, present three kinds of hyperbolic systems in the 
Ashtekar formulation of general relativity for Lorentzian vacuum 
spacetime.
We exhibit several (I) weakly hyperbolic, (II) diagonalizable
hyperbolic, and (III) symmetric hyperbolic systems, with each 
their eigenvalues. 
We demonstrate that Ashtekar's original equations form 
a weakly hyperbolic system. We discuss
how gauge conditions and reality conditions are constrained 
during each step toward constructing a symmetric hyperbolic system.

\end{abstract}



\section{Introduction} \label{sec:intro}

Developing hyperbolic formulations of the Einstein equation 
is growing into an important
research areas in general relativity \cite{Reula98}.
These formulations are used in the analytic proof of the existence, uniqueness
and stability (well-posedness)
of the solutions of the Einstein equation 
\cite{Heldbook}. 
So far, several first order hyperbolic formulations 
have been proposed;
some of them
are flux conservative \cite{BonaMasso},
some of them are symmetrizable 
or symmetric hyperbolic systems
\cite{FischerMarsden72,ChoquetBruhat,HF,fried96,VPE96,CBY,FR94,FR96}.
The recent interest in hyperbolic formulations
 arises from their application to numerical relativity.
One of the most useful features is the existence of  
characteristic speeds in hyperbolic systems.  
We expect more stable evolutions 
and expect imprements boundary conditions in their numerical 
simulation.
Some numerical tests have been reported along this direction
\cite{miguel,SBCSThyper,cactus1}.

Ashtekar's formulation of general relativity \cite{Ashtekar}
 has many
advantages.
By using his special pair of variables,
the constraint equations which appear in the theory become
low-order polynomials, and the theory has the
correct form for gauge theoretical interpretation. These features
suggest the possibility for developping a nonperturbative quantum
description of gravity.
Classical applications of the Ashtekar's formulation have also
been discussed by several authors. For
example, we \cite{ys-con} discussed
the reality conditions for the metric and triad and proposed a new
set of variables for Lorentzian dynamics.
We \cite{ysn-dege}
also showed an example of passing a degenerate
point in 3-space by locally relaxing the reality condition.
Although there is always a problem of reality
conditions in applying Ashtekar formulation to dynamics, 
we think that this new approach is quite attractive, and  broadens our
possibilities to attack dynamical issues.

A symmetric hyperbolic formulation of Ashtekar's variables 
was first developped by Iriondo, Leguizam\'on and Reula (ILR)
\cite{Iriondo}. 
They use {\it anti}-Hermiticity of
the principal symbol for defining their {\it symmetric} system.
Unfortunately, in their first short paper \cite{Iriondo}, 
they did not discuss the consistency of
their system with the reality conditions, which are
crucial in the study of the Lorentzian dynamics using
the Ashtekar variables.
We considered this point in \cite{YShypPRL}, and found that 
there are strict reality constraints 
(alternatively they can be interpreted as gauge conditions). 
Note that we primarily use the Hermiticity of the characteristic 
matrix to define a  symmetric hyperbolic system, which
we think the more conventional notation.
The difference between these definitions of symmetric hyperbolicity 
is commented in Appendix C.

The dynamical equations in the Ashtekar formulation of 
general relativity are themselves
quite close to providing a hyperbolic formulation. 
As we will show in \S \ref{sec4},
the original set of equations of motion is a 
first-order  (weakly) hyperbolic system.
One of the purposes of this paper is to develop
several hyperbolic systems based on the Ashtekar formulation
for Lorentzian vacuum spacetime, and discuss how gauge conditions 
and reality conditions are to be implemented. 
We categorize hyperbolic systems into three classes: (I) weakly 
hyperbolic (system has
all real eigenvalues), (II) diagonalizable hyperbolic
(characteristic matrix is diagonalizable), and (III)
symmetric hyperbolic system. These three classes have the relation
(III)  $\in$ (II) $\in$ (I), and 
are defined in detail in \S \ref{sec:def}. 
As far as we know, only a symmetric hyperbolic systems provide a fully 
well-posed  initial value formulation of partial 
differential equations systems.  However,  there are two reasons to 
consider the two other classes of hyperbolic systems, (I) and (II). 
First,  as we found in our
previous short paper \cite{YShypPRL}, the symmetric hyperbolic system 
we obtained using  Ashtekar's variables
has strict restrictions on the gauge conditions, while 
the original Ashtekar 
equations constitute a weakly hyperbolic system.  
We are interested in 
these differences, and show how additional constraints appear
during the steps toward constructing a symmetric hyperbolic system.
Second, 
many numerical experiments show that there are 
several advantages 
if we apply a certain form of hyperbolic formulation. 
Therefore, we think that presenting these three hyperbolic systems
is valuable to stimulate the studies in this field. 
To aid in possibly applying these systems
in numerical applications, 
we present characteristic speeds of each system we construct.

ILR, in their second paper \cite{ILRsecond},  expand their
previous discussion \cite{Iriondo} concerning 
reality conditions during evolution. 
They demand that the metric is
real-valued (metric reality condition),
and  use the freedom of internal rotation  
during the time evolution to set up
their soldering form so that it forms an anti-Hermitian 
principle symbol,
which is their basis to characterize the system symmetric. 
However, we adopt the view that re-defining inner product of the
fundamental variables 
introduces additional complications.
In our procedure, we first fix the inner product to construct
a symmetric hyperbolic system. 
As we will describe in \S \ref{sec5}, 
our symmetric
hyperbolic system then requires a reality condition on the triad
(triad reality condition), 
and in order to be consistent with its secondary condition
we need to impose further gauge conditions. 
The lack of these constraints in ILR, we believe, comes from their
incomplete treatment of a new gauge freedom, so-called 
{\it triad lapse} $\CA^a_0$ 
(discussed in \S \ref{sec:ash}), for dynamical evolutions in the 
Ashtekar formulation. 
In  Appendix \ref{appC}, 
we show that ILR's proposal
to use internal rotation to re-set triad reality does not work
if we adopt our conventional definition of hyperbolicity.

The layout of this paper is as follows:
In \S \ref{sec:def}, we define the three kinds of hyperbolic
systems which are considered in this paper. 
In \S \ref{sec:ash}, we briefly review 
Ashtekar's formulation and  the way of handling  reality conditions.
The following sections \S \ref{sec4} and \S \ref{sec5} are devoted to
constructing hyperbolic systems.  
Summary and discussion are in \S \ref{sec:disc}.
Appendix \ref{appA} supplements our proof of the uniqueness of our
symmetric hyperbolic system. 
Appendices \ref{appB} and \ref{appC} are  comments 
on ILR's treatment of the reality conditions. 

\section{Three definitions of hyperbolic systems} \label{sec:def}
\def\cha{J}
\def\coe{K}

We start by defining the hyperbolic
systems which are used in this paper.
\begin{Def}
We assume
a certain set of
(complex) variables $u_\alpha$ $(\alpha=1,\cdots, n)$
forms a first-order (quasi-linear)
partial differential equation system,
\begin{equation}
\partial_t u_\alpha
= \cha ^{l}{}^{\beta}{}_\alpha (u) \, \partial_l u_\beta
+\coe_\alpha(u),
\label{def}
\end{equation}
where $\cha$ (the characteristic matrix) and
$\coe$ are functions of $u$
but do not include any derivatives of $u$.
We say that the system (\ref{def}) is:
\been
\def\theenumi{(\Roman{enumi})}
\item  \label{weakhyp}
{\bf weakly hyperbolic},
if all the eigenvalues of the
characteristic matrix are real {\rm \cite{weakhyperbolic}}.
\item \label{diaghyp}
{\bf diagonalizable hyperbolic},
if the characteristic matrix is
diagonalizable and has all real eigenvalues
{\rm \cite{mizodiaghyp}}.
\item \label{symhyp}
{\bf symmetric hyperbolic},
if the characteristic matrix is a
Hermitian matrix {\rm \cite{fried96,CH}}.
\enen
\end{Def}

Here we state each definition more concretely.
We treat $\cha^{l\beta}{}_\alpha$ as a $n \times n$ matrix
when  the $l$-index is fixed.  
The following properties of these matrices are 
for every basis of $l$-index.

We say $\lambda^l$ is an
eigenvalue of $\cha^{l\beta}{}_\alpha$
when the characteristic equation,
$\det (\cha^{l\beta} {}_\alpha
-\lambda^l \delta^\beta {}_\alpha)=0$,
is satisfied.
The eigenvectors, $p^\alpha$, are given by solving
$\cha^{l}{}^{\alpha} {}_\beta \, p^{l\beta}=\lambda^l  p^{l\alpha}$.

The weakly hyperbolic system, \ref{weakhyp}, is
obtained when
 $\cha^l$ has {\it real spectrum} for every $l$, 
that is,  when
this characteristic equation can be divided
by $n$ real first-degree factors. 
{}For any single equation system,
the Cauchy problem under weak hyperbolicity is not,
 in general, $C^\infty$ well-posed, while it is solvable in the
class of the real analytic functions and in some
suitable Gevrey classes,
provided that the coefficients of the principal part
are sufficiently smooth.

The diagonalizable hyperbolic system, \ref{diaghyp},
is obtained when
$\cha$ is {\it real diagonalizable}, that is,
when
there exists complex regular matrix $P^l$
such that
$((P^l){}^{-1}) {}^\alpha {}_\gamma \, 
\cha^{l\gamma} {}_\delta  \, 
P^{l\delta} {}_\beta$
is real diagonal matrix for every $l$.
We can construct 
characteristic curves if the system is in this class. 
This system is often used as a model in the studies of
well-posedness in coupled linear hyperbolic system. 
(This is the same as {\it strongly hyperbolic} system 
as defined by some authors\cite{stewart,Gustaf}, but we use
the word {\it diagonalizable} since there exist other 
definitions for {\it strongly hyperbolic} 
systems\cite{taniguchisymp}.)

In order to define
the symmetric hyperbolic system, \ref{symhyp},
we need to declare an inner product
$\langle u|u \rangle$
to judge whether $\cha^{l\beta} {}_\alpha$ is Hermitian.
In other words, we are
required to define the way of lowering the index
$\alpha$ of $u^\alpha$.
We say $\cha^{l\beta} {}_\alpha$ is Hermitian
with respect to this index rule,
when
$\cha^l {}_{\beta\alpha}=\bar{\cha}^l {}_{\alpha\beta}$
 for every $l$,
where the overhead bar denotes complex conjugate.

Any Hermitian matrix is real diagonalizable, 
so that \ref{symhyp} $\in$ \ref{diaghyp}
$\in$ \ref{weakhyp}.
There are other
definitions of hyperbolicity; such as
{\it strictly hyperbolic} or {\it effectively hyperbolic},
if all eigenvalues of the characteristic
matrix are real and distinct (and non-zero for the latter).
These definitions are stronger than
\ref{diaghyp}, but exhibit no inclusion relation
with \ref{symhyp}.  In this paper, however, we only consider
\ref{weakhyp}-\ref{symhyp} above.

The symmetric system gives us the energy
integral inequalities,
which are the primary tools for studying
well-posedness of the system. 
As was discussed by Geroch \cite{Geroch},
most physical systems
can be expressed as symmetric hyperbolic systems.


\section{Ashtekar formulation}\label{sec:ash}
\subsection{Variables and Equations}
The key feature of  Ashtekar's formulation of general relativity
\cite{Ashtekar} is the introduction of a self-dual
connection as one of the basic dynamical variables.
Let us write 
the metric $g_{\mu\nu}$ using the tetrad 
$E^I_\mu$, with $E^I_\mu$ satisfying the gauge condition $E^0_a=0$.
Define its inverse, $E^\mu_I$, by
$g_{\mu\nu}=E^I_\mu E^J_\nu \eta_{IJ}$ and
$E^\mu_I:=E^J_\nu g^{\mu\nu}\eta_{IJ}$, 
where we use
$\mu,\nu=0,\cdots,3$ and
$i,j=1,\cdots,3$ as spacetime indices, while
$I,J=(0),\cdots,(3)$ and
$a,b=(1),\cdots,(3)$ are $SO(1,3)$, $SO(3)$ indices respectively.
We raise and lower
$\mu,\nu,\cdots$ by $g^{\mu\nu}$ and $g_{\mu\nu}$
(the Lorentzian metric);
$I,J,\cdots$ by $\eta^{IJ}={\rm diag}(-1,1,1,1)$ and $\eta_{IJ}$;
$i,j,\cdots$ by $\gamma^{ij}$ and $\gamma_{ij}$ (the 3-metric);
$a,b,\cdots$ by $\delta^{ab}$ and $\delta_{ab}$.
We also use volume forms $\epsilon_{abc}$: 
$\epsilon_{abc} \epsilon^{abc}=3!$.
We define SO(3,C) self-dual and anti self-dual
connections
\be
{}^{\pm\!}{\cal A}^a_{\mu}
:= \omega^{0a}_\mu \mp ({i / 2}) \epsilon^a{}_{bc} \, \omega^{bc}_\mu,
\en
where $\omega^{IJ}_{\mu}$ is a spin connection 1-form (Ricci
connection), $\omega^{IJ}_{\mu}:=E^{I\nu} \nabla_\mu E^J_\nu.$
Ashtekar's plan is to use  only a self-dual part of
the connection
$^{+\!}{\cal A}^a_\mu$
and to use its spatial part $^{+\!}{\cal A}^a_i$
as a dynamical variable.
Hereafter,
we simply denote $^{+\!}{\cal A}^a_\mu$ as ${\cal A}^a_\mu$.

The lapse function, $N$, and shift vector, $N^i$, both of which we
treat as real-valued functions,
are expressed as $E^\mu_0=(1/N, -N^i/N$).
This allows us to think of
$E^\mu_0$ as a normal vector field to $\Sigma$
spanned by the condition $t=x^0=$const.,
which plays the same role as that of Arnowitt-Deser-Misner (ADM) formulation.
Ashtekar  treated the set  ($\tilde{E}^i_{a}$, ${\cal A}^a_{i}$)
as basic dynamical variables, where
$\tilde{E}^i_{a}$ is an inverse of the densitized triad
defined by
\be
\tilde{E}^i_{a}:=e E^i_{a},
\en
where $e:=\det E^a_i$ is a density.
This pair forms the canonical set.

In the case of pure gravitational spacetime,
the Hilbert action takes the form
\begin{eqnarray}
S&=&\int {\rm d}^4 x
[ (\partial_t{\cal A}^a_{i}) \tilde{E}^i_{a}
+(i/2) \null \! \mathop {\vphantom {N}\smash N}
\limits ^{}_{^\sim}\!\null \tilde{E}^i_a
\tilde{E}^j_b F_{ij}^{c} \epsilon^{ab}{}_{c}
- 
\den^2
\Lambda \null \! \mathop {\vphantom {N}\smash N}
\limits ^{}_{^\sim}\!\null 
-N^i F^a_{ij} \tilde{E}^j_a
+{\cal A}^a_{0} \, {\cal D}_i \tilde{E}^i_{a} ],
 \label{action}
\end{eqnarray}
where 
$\null \! \mathop {\vphantom {N}\smash N}
\limits ^{}_{^\sim}\!\null := e^{-1}N$, 
${F}^a_{\mu\nu} := (d {\cal A}^a)_{\mu\nu}
-(i/2){\epsilon^a}_{bc}({\cal A}^b
 \wedge {\cal A}^c)_{\mu\nu}
= \partial_\mu {\cal A}^a_\nu
  - \partial_\nu {\cal A}^a_\mu
 - i \epsilon^{a}{}_{bc} \, {\cal A}^b_\mu{\cal A}^c_\nu
$
is the curvature 2-form,
$\Lambda$
is the cosmological constant,
${\cal D}_i \tilde{E}^j_{a}
    :=\partial_i \tilde{E}^j_{a}
-i \epsilon_{ab}{}^c  \, {\cal A}^b_{i}\tilde{E}^j_{c}$, 
 and
$\den^2=\det\tilde{E}^i_a 
=(\det E^a_i)^2$ is defined to be
$\det\tilde{E}^i_a=
(1/6)\epsilon^{abc}
\null\!\mathop{\vphantom {\epsilon}\smash \epsilon}
\limits ^{}_{^\sim}\!\null_{ijk}\tilde{E}^i_a \tilde{E}^j_b
\tilde{E}^k_c$, where
$\epsilon_{ijk}:=\epsilon_{abc}E^a_i E^b_j E^c_k$
 and $\null\!\mathop{\vphantom {\epsilon}\smash \epsilon}
\limits ^{}_{^\sim}\!\null_{ijk}:=e^{-1}\epsilon_{ijk}$
[When $(i,j,k)=(1,2,3)$,
we have
$\epsilon_{ijk}=e$,
$\null\!\mathop{\vphantom {\epsilon}\smash \epsilon}
\limits ^{}_{^\sim}\!\null_{ijk}=1$,
$\epsilon^{ijk}=\den^{-1}$, and 
$\tilde{\epsilon}^{ijk}=1$.].

Varying the action with respect to the non-dynamical variables
$\null \!
\mathop {\vphantom {N}\smash N}\limits ^{}_{^\sim}\!\null$,
$N^i$
and ${\cal A}^a_{0}$ yields the constraint equations,
\begin{eqnarray}
{\cal C}_{H} &:=&
 (i/2)\epsilon^{ab}{}_c \, 
\tilde{E}^i_{a} \tilde{E}^j_{b} F_{ij}^{c}
  -\Lambda \, \det\tilde{E}
   \approx 0, \label{c-ham} \\
{\cal C}_{M i} &:=&
  -F^a_{ij} \tilde{E}^j_{a} \approx 0, \label{c-mom}\\
{\cal C}_{Ga} &:=&  {\cal D}_i \tilde{E}^i_{a}
 \approx 0.  \label{c-g}
\end{eqnarray}
The equations of motion for the dynamical variables
($\tilde{E}^i_a$ and ${\cal A}^a_i$) are
\begin{eqnarray} 
\partial_t {\tilde{E}^i_a}
&=&-i{\cal D}_j( \epsilon^{cb}{}_a  \, \null \!
\mathop {\vphantom {N}\smash N}\limits ^{}_{^\sim}\!\null
\tilde{E}^j_{c}
\tilde{E}^i_{b})
+2{\cal D}_j(N^{[j}\tilde{E}^{i]}_{a})
+i{\cal A}^b_{0} \epsilon_{ab}{}^c  \, \tilde{E}^i_c,  \label{eqE}
\\
\partial_t {\cal A}^a_{i} &=&
-i \epsilon^{ab}{}_c  \, 
\null \! \mathop {\vphantom {N}\smash N}
\limits ^{}_{^\sim}\!\null \tilde{E}^j_{b} F_{ij}^{c}
+N^j F^a_{ji} +{\cal D}_i{\cal A}^a_{0}+\Lambda \null \!
\mathop {\vphantom {N}\smash N}\limits ^{}_{^\sim}\!\null 
\tilde{E}^a_i,
\label{eqA} 
\end{eqnarray}
\noindent
where
${\cal D}_jX^{ji}_a:=\partial_jX^{ji}_a-i
 \epsilon_{ab}{}^c {\cal A}^b_{j}X^{ji}_c,$
 for $X^{ij}_a+X^{ji}_a=0$.

In order to construct metric variables from the variables
$(\tilde{E}^i_a, {\cal A}^a_i,  \null \!
\mathop {\vphantom {N}\smash N}\limits ^{}_{^\sim}\!\null, N^i)$,
we first prepare
tetrad $E^\mu_I$ as
$E^\mu_{0}=({1 / e \null \! \mathop {\vphantom {N}\smash N}
\limits ^{}_{^\sim}\!\null}, -{N^i / e \null \!
\mathop {\vphantom {N}\smash N}\limits ^{}_{^\sim}\!\null})$ and
$E^\mu_{a}=(0, \tilde{E}^i_{a} /e).$
Using them, we obtain metric $g^{\mu\nu}$ such that
\begin{equation}
g^{\mu\nu}:=E^\mu_{I} E^\nu_{J} \eta^{IJ}. \label{recmet}
\end{equation}

\subsection{Reality conditions}
The metric (\ref{recmet}), in general, is not real-valued in the Ashtekar
formulation.
To ensure that the metric is real-valued,
we need to impose real lapse and shift vectors together with
two conditions (the metric reality condition);
\begin{eqnarray}
\im (\tilde{E}^i_a \tilde{E}^{ja} ) &=& 0, \label{w-reality1} \\
\re (\epsilon^{abc}
\tilde{E}^k_a \tilde{E}^{(i}_b {\cal D}_k \tilde{E}^{j)}_c)
&=& 0,
\label{w-reality2-final}
\end{eqnarray}
where the latter comes from the secondary condition of reality
of the metric
$\im \{ \partial_t(\tilde{E}^i_a \tilde{E}^{ja} ) \} = 0$
\cite{AshtekarRomanoTate}, and
we assume
$\det\tilde{E}>0$ (see \cite{ys-con}).
These metric reality conditions, 
(\ref{w-reality1}) and 
(\ref{w-reality2-final}), are automatically preserved during the evolution
 if the variables satisfy the conditions on the initial data 
 \cite{AshtekarRomanoTate,ys-con}.

For later convenience, we also prepare
stronger reality conditions.
These conditions are
\begin{eqnarray}
\im (\tilde{E}^i_a ) &=& 0
\label{s-reality1} \\
{\rm and~~}
\im  ( \partial_t {\tilde{E}^i_a} ) &=& 0,
\label{s-reality2}
\end{eqnarray}
\noindent
and we call them the ``primary triad reality condition" and the
``secondary triad
reality condition", respectively.
Using the equations of motion of $\tilde{E}^i_{a}$,
the gauge constraint (\ref{c-ham})-(\ref{c-g}),
the metric reality conditions
(\ref{w-reality1}), (\ref{w-reality2-final})
and the primary condition (\ref{s-reality1}),
we see  that  (\ref{s-reality2}) is equivalent to \cite{ys-con}
\begin{equation}
\re({\cal A}^a_{0})=
\partial_i( \null \! \mathop {\vphantom {N}\smash N}
\limits ^{}_{^\sim}\!\null )\tilde{E}^{ia}
+(1 /2e) E^b_i \null \! \mathop {\vphantom {N}\smash N}
\limits ^{}_{^\sim}\!\null
\tilde{E}^{ja} \partial_j\tilde{E}^i_b
+N^{i} \re({\cal A}^a_i), \label{s-reality2-final}
\end{equation}
or with un-densitized variables,
\begin{equation}
\re({\cal A}^a_{0})=
\partial_i( N)
{E}^{ia}
+N^{i} \, \re({\cal A}^a_i).
\label{s-reality2-final2}
\end{equation} 
{}From this expression we see that
the secondary triad reality condition
restricts the three components of ``triad lapse" vector
${\cal A}^a_{0}$.
Therefore (\ref{s-reality2-final}) is
not a restriction on the dynamical variables
($\tilde{E}^i_a $ and ${\cal A}^a_i$)
but on the slicing, which we should impose on each hypersurface.
Thus the secondary triad reality condition does not restrict the
dynamical variables any
further than the secondary metric condition does.

Throughout this paper, we basically impose metric
reality condition.  We assume that initial data of
$(\dtri^i_a, \CA^a_i)$ for evolution are solved so as
to satisfy all three constraint equations and metric
reality condition (\ref{w-reality1}) and (\ref{w-reality2-final}).
Practically, this is
obtained, for
example, by solving ADM constraints and by transforming a
set of initial data to Ashtekar's notation.

\subsection{Characteristic matrix}
We shall see how the
definitions of hyperbolic systems in \S \ref{sec:def}
 can be applied for
Ashtekar's equations of motion
(\ref{eqE}) and (\ref{eqA}).
Since both dynamical variables,
$\dtri^i_a$ and $\CA^a_i$, have 9 components each
(spatial index: $i=1,2,3$ and
SO(3) index: $a=(1),(2),(3)$),
the combined set of variables, $u^\alpha =
(\dtri^i_a, \CA^a_i) $,  has 18 components.
Ashtekar's formulation itself is
in the first-order  (quasi-linear)
form in the sense of (\ref{def}),
but is not in a symmetric hyperbolic form.

We start by writing the
principal part of the Ashtekar's evolution equations as
\begin{equation}
\partial_t \left[ \begin{array}{l}
\tilde{E}^i_a \\
\CA^a_i
\end{array} \right] \cong
\left[ \begin{array}{cc}
A^l {}_a {}^{bi}{}_j & 
B^l{}_{ab}{}^{ij} \\
C^{lab}{}_{ij} & 
D^{la}{}_{bi}{}^j
\end{array} \right]
\partial_l
\left[ \begin{array}{l}
\tilde{E}^j_b \\
\CA^b_j
\end{array} \right],
\label{matrixform}
\end{equation}
where
$\cong$ means that we have extracted only the terms which
appear in the principal part of the system. We name
these components as $A,B,C$ and $D$ for later convenience.

The characteristic equation  becomes
\be
\det \left( \begin{array}{cc}
A^l {}_a {}^{bi}{}_j 
-\lambda^l \delta^b_a\delta^i_j & 
B^l{}_{ab}{}^{ij} \\
C^{lab}{}_{ij} & 
D^{la}{}_{bi}{}^j
-\lambda^l\delta^a_b\delta^j_i
\end{array} \right)=0.
\en
If $B^l{}_{ab}{}^{ij}$ and
$C^{lab}{}_{ij}$ vanish, then 
the characteristic matrix is diagonalizable
if $A$ and $D$ are diagonalizable, since the spectrum of the
characteristic matrix is composed of those of $A$ and $D$.
The eigenvectors for every $l$-index, $(p^l{}^i_a, q^l {}^a_i)$, are given by
\be
 \left( \begin{array}{cc}  
A^l {}_a {}^{bi}{}_j & 
B^l{}_{ab}{}^{ij} \\
C^{lab}{}_{ij} & 
D^{la}{}_{bi}{}^j
\end{array} \right)
\left( \matrix{p^l {}^j_b \cr q^l {}^b_j }\right)
=\lambda^l
\left( \matrix{p^l {}^i_a \cr q^l {}^a_i }\right)
\mbox{ for every }l.
\en

The lowering rule for the $\alpha$ of $u^\alpha$
follows those of the spacetime or internal indices.
The corresponding inner product takes the form
$\langle u| u \rangle :=u_\alpha \bar{u}^\alpha$.
According to this rule,
we say the characteristic matrix is a Hermitian when
\begin{eqnarray}
0&=&
A^{labij}-\bar{A}^{lbaji},  \label{cond1}
\\
0&=&
D^{labij}
-\bar{D}^{lbaji}, \label{cond2}
\\
0 &=&
B^{labij}-\bar{C}^{lbaji}. \label{cond3}
\end{eqnarray}

\section{Constructing
hyperbolic systems with original equations of motion}
\label{sec4}
In this section, we consider which form of hyperbolicity applies to
the original equations of motion,
(\ref{eqE}) and (\ref{eqA}),
under the metric reality condition (\S \ref{sec41}) or
under the triad reality condition (\S \ref{sec42}).

\subsection{
Under metric reality condition (system Ia and IIa)}
\label{sec41}
As the first approach, we take the
equations of motion
(\ref{eqE}) and (\ref{eqA}) with metric reality conditions
(\ref{w-reality1}) and (\ref{w-reality2-final}).
The principal term of (\ref{eqE}) and (\ref{eqA})
become
\beas
\partial_t \tilde{E}^i_a
&=&
-i{\cal D}_j( \epsilon^{cb}{}_{a} \,
\null \! \mathop {\vphantom {N}\smash N}
\limits ^{}_{^\sim}\!\null
\tilde{E}^j_c\tilde{E}^i_b)
+2{\cal D}_j(N^{[j}\tilde{E}^{i]}_a)
+i{\cal A}^b_0 \epsilon_{ab}{}^{c}  \, \tilde{E}^i_c
\nonumber\\&\cong&
-i\epsilon^{cb}{}_{a} \, 
\null \! \mathop {\vphantom {N}\smash N}
\limits ^{}_{^\sim}\!\null
(\partial_j\tilde{E}^j_c)\tilde{E}^i_b
-i\epsilon^{cb}{}_{a} \, 
\null \! \mathop {\vphantom {N}\smash N}
\limits ^{}_{^\sim}\!\null
\tilde{E}^j_c(\partial_j\tilde{E}^i_b)
+{\cal D}_j(N^j\tilde{E}^i_a)
-{\cal D}_j(N^i\tilde{E}^j_a)
\nonumber\\&\cong&
[-i\epsilon^{bc}{}_{a} \, 
\null \! \mathop {\vphantom {N}\smash N}
\limits ^{}_{^\sim}\!\null
\delta^l_j  \tilde{E}^i_c
-i\epsilon^{cb}{}_{a} \, 
\null \! \mathop {\vphantom {N}\smash N}
\limits ^{}_{^\sim}\!\null
\tilde{E}^l_c \delta^i_j
+N^l\delta^i_j \delta^b_a
-N^i\delta^l_j \delta^b_a ]
(\partial_l\tilde{E}^j_b),
\\
\partial_t {\cal A}^a_i &=&
-i \epsilon^{ab}{}_{c} \, 
\null \! \mathop {\vphantom {N}\smash N}
\limits ^{}_{^\sim}\!\null
\tilde{E}^j_b F^c_{ij}
+N^j F^a_{ji}
+{\cal D}_i{\cal A}^a_0+\Lambda \,
\null \! \mathop {\vphantom {N}\smash N}
\limits ^{}_{^\sim}\!\null
\tilde{E}^a_i
\nonumber\\&\cong&
-i \epsilon^{ab}{}_{c} \, 
\null \! \mathop {\vphantom {N}\smash N}
\limits ^{}_{^\sim}\!\null
\tilde{E}^j_b (\partial_i{\cal A}^c_j
-\partial_j{\cal A}^c_i)
+N^j (\partial_j{\cal A}^a_i-\partial_i{\cal A}^a_j)
\nonumber\\&\cong&
[
+i \epsilon^a{}_b {}^c \, 
\null \! \mathop {\vphantom {N}\smash N}
\limits ^{}_{^\sim}\!\null
\tilde{E}^j_c \delta^l_i
-i \epsilon^a{}_b{}^c \, 
\null \! \mathop {\vphantom {N}\smash N}
\limits ^{}_{^\sim}\!\null
\tilde{E}^l_c \delta^j_i
+N^l \delta^a_b \delta^j_i
-N^j \delta^a_b \delta^l_i
](\partial_l{\cal A}^b_j).
\enas
The principal terms in the notation of (\ref{matrixform}) become
\bear
A^{labij}&=&
-i\ut N \ep^{abc} \dtri^i_c \gamma^{lj}
+i\ut N \ep^{abc} \dtri^l_c \gamma^{ij}
+N^l \delta^{ab} \gamma^{ij}
-N^i\delta^{ab} \gamma^{lj},
\label{ori-A}
\\
B^{labij}&=& C^{labij}=0,
\label{ori-BC}
\\
D^{labij}&=&
+i\ut N \ep^{abc} \dtri^j_c \gamma^{li}
-i\ut N \ep^{abc} \dtri^l_c \gamma^{ij}
+N^l \delta^{ab} \gamma^{ij}
-N^j \delta^{ab} \gamma^{li}.
\label{ori-D}
\enar
We get the 18 eigenvalues of the characteristic matrix,
all of which are independent of the choice of  triad:
\beas
0 ~(\mbox{multiplicity=} 6), ~
N^l ~(4), ~
N^l \pm N\sqrt{\gamma^{ll}} ~(4 \mbox{ each}),
\enas
where we do {\it not} take the sum in $\gamma^{ll}$
(and we maintain this notation 
hereafter for eigenvalues and related discussions).
Therefore we can say that
this system is weakly hyperbolic, of type \ref{weakhyp}.

We note that this system is not type \ref{diaghyp} in general,
because this is not diagonalizable, for example, when
$N^l=0$.
We classify this system as type \ref{weakhyp},
and call this {\it system Ia}, hereafter.

The necessary and sufficient
conditions to make
this system diagonalizable, type \ref{diaghyp}, are that
the gauge conditions
\be
N^l \neq 0
\mbox{~nor~}
\pm N \sqrt{\gamma^{ll}}
,
\qquad
\mbox{and~}
\gamma^{ll} >0
\label{IIa_1}
\en
where the last one is the positive definiteness of $\gamma^{ll}$.
This can be proved as follows.
Suppose that $(\ref{IIa_1})$ is satisfied.
Then $0,N^l,N^l\pm \sqrt{\gamma^{ll}}$
are four distinct eigenvalues
and we see
$\rank(\cha^l)=12$,
$\rank(\cha^l-N^l I)=14$,
$\rank(\cha^l-(N^l\pm N \sqrt{\gamma^{ll}})I)=14$.
Therefore the characteristic matrix is diagonalizable.
Conversely suppose that
$N^l=0$ or
$N^l=\pm N \sqrt{\gamma^{ll}}$, then we see
the characteristic matrix is not diagonalizable
in each case.

The components of the characteristic matrix are
the same as {\it system Ia},
so all eigenvalues are equivalent with {\it system Ia}.
We can also show that this system is not Hermitian
hyperbolic.
Therefore we classify the system
[{\it Ia} + (\ref{IIa_1})] to real
diagonalizable hyperbolic, type \ref{diaghyp}, and
call this set as {\it system IIa}.
However, we will show
in the next section that real diagonalizable hyperbolic
system can also be constructed with less strict gauge
conditions by modifying right-hand-side of equations of
motion ({\it system IIb}).

\subsection{Under triad reality condition (system Ib)}\label{sec42}
Next, we consider systems of the original
equations of motion,
(\ref{eqE}) and (\ref{eqA}), with the triad reality
condition.
Since
this reality condition
requires the additional
 (\ref{s-reality2-final}) or (\ref{s-reality2-final2})
as the secondary condition (that is, to preserve the reality of
triad during time evolution),
in order to be consistent with this requirement and to
avoid the system becoming second order in
fundamental variables, we need to set
$\ptl_iN=0$. This fixes the real part of the {\it triad lapse} gauge as
$\re(\CA^a_0)=\re(\CA^a_iN^i)$.
We naturally define its imaginary part as
$\im(\CA^a_0)=\im(\CA^a_iN^i)$.
Thus the triad lapse is fixed as
$\CA^a_0=\CA^a_iN^i$.
This gauge restriction does not affect principal
part of the evolution equation
for $\dtri^i_a$, but requires us to add the term
\beas
{\cal D}_i{\cal A}^a_0
&\cong&
\ptl_i{\cal A}^a_0
=
\ptl_i(\CA^a_jN^j)
\cong
N^j(\ptl_i\CA^a_j)
=
N^j\delta^a_b\delta^l_i(\ptl_l\CA^b_j)
\enas
to the right-hand-side of the equation of $\CA^a_i$.
That is,
we need to add
$N^j\delta^a_b\delta^l_i$
to $D^{la}{}_{bi}{}^j$ in (\ref{ori-D}),
\beas
D^{labij}
&=&
 i \epsilon^{abc} \ut N \tilde{E}^j_c \gamma^{li}
-i \epsilon^{abc} \ut N \tilde{E}^l_c \gamma^{ji}
+N^l \delta^{ab} \gamma^{ji}.
\enas
The other components of the characteristic matrix
remain the same [(\ref{ori-A}) and (\ref{ori-BC})].
We find that the set of eigenvalues of this system is
\beas
0 \ (\mbox{multiplicity =}3), \qquad
N^l \ (7), \qquad
N^l\pm N \sqrt{\gamma^{ll}}\ (4 \mbox{ each}).
\enas
Therefore the system is again, type \ref{weakhyp}.
This system is not
real diagonalizable because $D^l$ is not.
So we classify this system as type \ref{weakhyp}
and call this set as {\it system Ib}.
We note that
this system is not
real diagonalizable
for any choice of gauge.
Therefore we cannot construct a system of type \ref{diaghyp}
using the same technique of constructing {\it system IIa}.
However, as we will show in the next section,
the system becomes diagonalizable (and symmetric) hyperbolic
under the triad reality condition
if we modify the equations of motion.

\section{Constructing a symmetric hyperbolic system}
\label{sec5}
{}From the analysis of the previous section, we found that the
original set of
equations of motion in the Ashtekar formulation constitute a
weakly hyperbolic system, type \ref{weakhyp}, 
or a diagonalizable hyperbolic system, type \ref{diaghyp},
under appropriate gauge conditions, but we
also found that we could not obtain a symmetric hyperbolic system, 
type \ref{symhyp}.
In this section, we show that type \ref{symhyp} is obtained 
if we modify the equations of motion.
We begin by describing our approach without considering reality
conditions, but we will soon show that the triad reality
condition is required for making the characteristic matrix
Hermitian.

We first prepare the constraints
(\ref{c-ham})-(\ref{c-g})  as
\begin{eqnarray}
{\cal C}_{H} &\cong&
 i\epsilon^{ab}{}_{c}  \, \tilde{E}^i_a \tilde{E}^j_b
\partial_i{\cal A}^c_j
=
 i\epsilon^{dc}{}_{b}  \, \tilde{E}^l_d \tilde{E}^j_c
(\partial_l{\cal A}^b_j)
=
 -i\epsilon_b{}^{cd}  \, \tilde{E}^j_c \tilde{E}^l_d
(\partial_l{\cal A}^b_j),
\\
{\cal C}_{M k} &=&
  -F^a_{kj} \tilde{E}^j_a
\cong
  -(\partial_k{\cal A}^a_j-\partial_j {\cal A}^a_k) \tilde{E}^j_a
=
[-\delta^l_k \tilde{E}^j_b
+\delta^j_k \tilde{E}^l_b
](\partial_l{\cal A}^b_j),
\\
{\cal C}_{Ga} &=&
 {\cal D}_i \tilde{E}^i_a
\cong
\partial_l\tilde{E}^l_a.
\end{eqnarray}
We apply the same technique as used by ILR to modify the
equation of motion of $\tilde{E}^i_a$ and ${\cal A}^a_i$; by
adding the constraints which weakly produce
${\cal C}_{H} \approx 0, \,
{\cal C}_{M k} \approx 0,$  and
${\cal C}_{Ga} \approx 0$. (Indeed, this technique has also been used
for constructing symmetric hyperbolic systems for the original 
Einstein equations \cite{CBY,FR96}.)
We also assume the triad lapse ${\cal A}^a_0$ is
\begin{equation}
\partial_i{\cal A}^a_0 \cong
 T^{lab} {}_{ij}  \, \partial_l \tilde{E}^j_b
+S^{la} {}_{bi} {}^j  \, \partial_l {\cal A}^b_j, \label{A0katei}
\end{equation}
where $T$ and $S$ are parameters which do not include derivatives
of the fundamental variables. This assumption is general for our
purpose of studying the principal part of the system.

One natural way to construct a symmetric hyperbolic system is
to keep $B=C=0$ and modify the $A$ and $D$ terms in (\ref{matrixform}),
so that we modify (\ref{eqE}) using ${\cal C}_{G}$, and
modify (\ref{eqA})  using ${\cal C}_{H}$ and  ${\cal C}_{M}$.
That is, we add the following terms to the equations of motion:
\begin{eqnarray}
\mbox{modifying term for }\partial_t \tilde{E}^i_a 
&=&
P^i{}_{ab}  \, {\cal C}_G^b
\cong
P^i{}_a{}^b  \, \partial_j\tilde{E}^j_b
=
(P^i{}_a{}^b  \, \delta^l_j)(\partial_l\tilde{E}^j_b),
\label{eqE2}
\\
\mbox{modifying term for }
\partial_t {\cal A}^a_i 
&=&
{\cal D}_i{\cal A}^a_0
+Q^a_i {\cal C}_H
+R_i{}^{ja}  \, {\cal C}_{Mj}
\nonumber \\ &\cong&
 T^{lab} {}_{ij}  \, \partial_l \tilde{E}^j_b
+S^{la} {}_{bi} {}^j  \, \partial_l {\cal A}^b_j
-i Q^a_i \epsilon_b{}^{cd} \, 
\tilde{E}^j_c \tilde{E}^l_d  (\partial_l{\cal A}^b_j)
+R_i{}^{ka} [-\delta^l_k \tilde{E}^j_b
+\delta^j_k \tilde{E}^l_b ]
\partial_l{\cal A}^b_j
\nonumber\\&\cong&
[
 S^{la} {}_{bi} {}^j
-iQ^a_i \epsilon_b{}^{cd}  \, \tilde{E}^j_c \tilde{E}^l_d
-R_i{}^{la}  \, \tilde{E}^j_b
+R_i{}^{ja}  \, \tilde{E}^l_b
](\partial_l{\cal A}^b_j)
+T^{lab} {}_{ij}  \, \partial_l \tilde{E}^j_b, \label{eqA2}
\end{eqnarray}
where $P, Q$, and $R$ are parameters and will be
fixed later.
In Appendix \ref{appA2}, we show that the modifications to
the off-diagonal blocks $B$ and $C$, i.e.
modifying (\ref{eqE}) using ${\cal C}_{H}$  and  ${\cal C}_{M}$ and
modify (\ref{eqA})  using ${\cal C}_{G}$, 
will not affect the final conclusion at all. 
Note that
we truncated ${\cal A}^a_0$ in (\ref{eqE2}),
while it remains in  (\ref{eqA2}), since
only the derivative of  ${\cal A}^a_0$ effects the principal
part of the system. The terms in (\ref{matrixform}) become
\begin{eqnarray}
A^{labij}&=&
-i\epsilon^{bca}
\null \! \mathop {\vphantom {N}\smash N}\limits ^{}_{^\sim}\!\null
\gamma^{lj}\tilde{E}^i_c
-i\epsilon^{cba}
\null \! \mathop {\vphantom {N}\smash N}\limits ^{}_{^\sim}\!\null
\tilde{E}^l_c \gamma^{ij}
+N^l\gamma^{ij} \delta^{ab}
-N^i\gamma^{lj} \delta^{ab}
+P^{iab} \gamma^{lj},
\\
B^{labij}&=&0,
\\
C^{labij}&=&
T^{labij},
\\
D^{labij}&=&
 i \epsilon^{abc}
\null \! \mathop {\vphantom {N}\smash N}\limits ^{}_{^\sim}\!\null
\tilde{E}^j_c \gamma^{li}
-i \epsilon^{abc}
\null \! \mathop {\vphantom {N}\smash N}\limits ^{}_{^\sim}\!\null
\tilde{E}^l_c \gamma^{ji}
+N^l \delta^{ab} \gamma^{ji}
-N^j \delta^{ab} \gamma^{li}
+S^{labij}
\nonumber \\ &&
-iQ^{ai} \epsilon^{bcd} \tilde{E}^j_c \tilde{E}^l_d
-R^{ila} \tilde{E}^{jb}
+R^{ija} \tilde{E}^{lb}. \label{D}
\end{eqnarray}

The condition (\ref{cond3}) immediately shows $T^{labij}=0$.
The condition (\ref{cond1}) is
written as
\begin{eqnarray}
0&=&
-i\ep^{abc} \ut N \gamma^{lj} \dtri^i_c
+i\ep^{abc} \ut N \gamma^{li} \bar{\dtri}{}^j_c
-2\ep^{abc} \ut N \gamma^{ij} \, \im(\dtri^l_c)
\nonumber \\&&
-N^i\gamma^{lj} \delta^{ab}
+N^j\gamma^{li} \delta^{ab}
+P^{iab} \gamma^{lj}
-\bar{P}^{jba} \gamma^{li} := \dagger^{labij}. \label{a-a}
\end{eqnarray}
By contracting $\dagger^l{}_{ab}{}^{ij}$,
we get
$\re(\ep_{abc}\dagger^{labik}\gamma_{li}
-2\ep_{abc}\dagger^{kabij}\gamma_{ij})
=20 \, \ut N \, \im(\dtri^k_c)$.
This suggests that we should impose $ \im( \tilde{E}^l_c) =0$,
in order to get $\dagger^l{}_{ab}{}^{ij}=0$.
This means that the triad reality condition is required
for making the characteristic matrix Hermitian.

\subsection{Under triad reality condition (system IIIa)}
\label{sym}
In this subsection, 
we assume the triad reality condition hereafter. 
In order to be consistent with the secondary triad reality condition
(\ref{s-reality2-final2}) during time evolution, and in order to
avoid the system becoming second order, we
need to specify the lapse function as $\partial_i N=0$.
This lapse condition reduces to 
\bear
\re(\CA^a_0)&=&
N^{i} \, \re({\cal A}^a_i),
\label{A0}
\\
\ptl_i \, \re(\CA^a_0)&\cong&
N^j\ptl_i \, \re({\cal A}^a_j).
\label{A0bibun}
\enar
By comparing these with the  real and 
imaginary components of (\ref{A0katei}), i.e., 
\bear
\partial_i \, \re({\cal A}^a_0) &\cong&
 \re(S^{la} {}_{bi} {}^j) \, \partial_l  \, \re({\cal A}^b_j)
-\im(S^{la} {}_{bi} {}^j) \, \partial_l  \, \im({\cal A}^b_j),
\label{A0re}
\\
\partial_i \, \im({\cal A}^a_0) &\cong&
 \im(S^{la} {}_{bi} {}^j) \, \partial_l  \, \re({\cal A}^b_j)
+\re(S^{la} {}_{bi} {}^j) \, \partial_l  \, \im({\cal A}^b_j),
\label{A0im}
\enar
we obtain
\beas
\re(S^{la} {}_{bi} {}^j)=N^j \delta^l_i \, \delta^{a}_b
\qquad \mbox{\rm and} \qquad
\im(S^{la} {}_{bi} {}^j)=0.
\enas
Thus $S$ is determined as
\be
S^{la} {}_{bi} {}^j=N^j \delta^l_i \, \delta^{a}_b.
\label{S}
\en

This value of $S$,  and ${T}=0$, 
determine the form of the triad lapse as
\begin{equation}
{\cal A}^a_0 = {\cal A}^a_i N^i+\mbox{non-dynamical~terms}.
\label{shift-triad-relation}
\end{equation}
ILR do not discuss consistency of their system with the 
reality condition (especially with the secondary reality condition).
However, since ILR assume ${\cal A}^a_0 = {\cal A}^a_iN^i$,
we think that ILR also need to impose a similar restricted lapse
condition in order to preserve reality of their system.

By decomposing  $\dagger$, that is (\ref{cond1}),  into its
real and imaginary parts, we
get
\beas
0
&=&
-N^i\gamma^{lj} \delta^{ab}
+N^j\gamma^{li} \delta^{ba}
+ \gamma^{lj} \, \re(P)^{iab}
- \gamma^{li} \, \re(P)^{jba},
\\
0
&=&
-\epsilon^{bca} \ut N
\gamma^{lj}\tilde{E}^i_c
-\epsilon^{acb} \ut N
\gamma^{li}\tilde{E}^j_c
+ \gamma^{lj} \, \im(P)^{iab}
+ \gamma^{li} \, \im(P)^{jba}.
\enas
By multiplying $\gamma_{li}$ to these and
taking symmetric and anti-symmetric components on the indices $ab$,
we have
\beas
0
&=&
 2 N^j\delta^{(ba)}
+\re(P)^{j(ab)}
-3\re(P)^{j(ba)}
=
 2 N^j\delta^{ba}
-2\re(P)^{j(ab)},
\\
0
&=&
 2 N^j\delta^{[ba]}
+\re(P)^{j[ab]}
-3\re(P)^{j[ba]}
=
4\re(P)^{j[ab]},
\\
0
&=&
 \im(P)^{j(ab)}
+3\im(P)^{j(ba)}
=
4\im(P)^{j(ab)},
\\
0
&=&
-2\epsilon^{acb}\ut N \tilde{E}^j_c
+\im(P)^{j[ab]}
+3\im(P)^{j[ba]}
=
-2\epsilon^{acb}\ut N \tilde{E}^j_c
-2\im(P)^{j[ab]}.
\enas
These imply
\begin{equation}
P^{iab}=
N^i \delta^{ab}+i\ut N \epsilon^{abc}\tilde{E}^i_c.
\label{paraP}
\end{equation}

Our task is finished when we specify the  parameters
$Q$ and $R$.
By substituting (\ref{S}) into (\ref{D}),
the condition (\ref{cond2})
 becomes
\begin{eqnarray}
0
&=&
i \epsilon^{abc} \ut N \dtri^j_c \gamma^{li}
+i \epsilon^{bac} \ut N \dtri^i_c \gamma^{lj}
-iQ^{ai} \epsilon^{bcd} \tilde{E}^j_c \tilde{E}^l_d
-i\bar{Q}^{bj} \epsilon^{acd} \tilde{E}^i_c \tilde{E}^l_d
-R^{ila} \tilde{E}^{jb}
+R^{ija} \tilde{E}^{lb} \nonumber \\ &&
+\bar{R}^{jlb} \tilde{E}^{ia}
-\bar{R}^{jib} \tilde{E}^{la}.
\label{d-d}
\end{eqnarray}
We found that a combination of the choice
\begin{eqnarray}
Q^{ai}&=&
\den^{-2}
\null \! \mathop {\vphantom {N}\smash N}\limits ^{}_{^\sim}\!\null
\tilde{E}^{ia}, 
\mbox{~and~}
R^{ila}=
i \den^{-2}
\null \! \mathop {\vphantom {N}\smash N}\limits ^{}_{^\sim}\!\null
\epsilon^{acd} \tilde{E}^i_d \tilde{E}^l_c,
\label{paraQR}
\end{eqnarray}
satisfies the condition (\ref{d-d}).  
We show in Appendix \ref{appA1}
that this pair of
$Q$ and $R$ satisfies (\ref{d-d})
and that this choice is unique.

The final equations of motion are
\bear
A^{labij}&=&
i\ep^{abc}  \ut N
\dtri^l_c \gamma^{ij}
+N^l\gamma^{ij} \delta^{ab},
\label{fm-A}
\\
B^{labij}&=&C^{labij}=0,
\label{fm-BC}
\\
D^{labij}&=&
i\ut N(\ep^{abc} \dtri^j_c \gamma^{li}
- \ep^{abc} \dtri^l_c \gamma^{ji} \nonumber \\ &~&
-\den^{-2} \dtri^{ia} \ep^{bcd} \dtri^j_c \dtri^l_d
-\den^{-2}\ep^{acd} \dtri^i_d \dtri^l_c  \dtri^{jb}
+\den^{-2}
\ep^{acd} \dtri^i_d \dtri^j_c \dtri^{lb}
)
+N^l \delta^{ab} \gamma^{ij}.
\label{fm-D}
\enar
To summarize, we obtain a symmetric hyperbolic system, 
type \ref{symhyp}
by modifying the equations of motion, 
restricting the gauge to: $\CA^a_0=\CA^a_iN^i$, $\ptl_i N=0$, 
and assuming the triad reality condition.
We name this set {\it system IIIa}.
The eigenvalues of this system are 
\be
N^l ~\mbox{(multiplicity = 6)}, \qquad
N^l \pm \sqrt{\gamma^{ll}} N ~\mbox{(5~each)} \qquad
\mbox{and} \qquad
N^l \pm \sqrt{\gamma^{ll}} N~\mbox{(1~each)}.
\label{symhypIIIa_eigen}
\en
These speeds are again independent of the choice of (real) triad.

\subsection{Under metric reality condition (system IIb)}

Using this technique, we can also construct another example of 
diagonalizable hyperbolic system.
Since the parameters $S$ and $T$ specify {\it triad
lapse}, a gauge variable for time evolutions,
it is possible to change our interpretation
that we take the evolution of the system within the 
{\it metric} reality condition.
Of course, the characteristic matrix is no longer  Hermitian.
{}From the fact that
we do not use the triad reality condition in the
process of modifying the characteristic matrix using parameters
(\ref{paraP}), (\ref{paraQR})
nor in the process of
deriving the eigenvalues, this system has the same components
in its characteristic matrix and has the same eigenvalues.
The process of examining diagonalizability is 
independent of the reality conditions. Therefore
this system is classified as a diagonalizable hyperbolic
system, type \ref{diaghyp}.

To summarize, 
we gain another diagonalizable hyperbolic system
by modifying the equations of motion using terms from constraint
equations, with characteristic matrix (\ref{fm-A})-(\ref{fm-D})
under metric reality condition. The eigenvalues are
(\ref{symhypIIIa_eigen}), and this system is restricted only by
a condition on triad lapse, $\CA^a_0=\CA^a_iN^i$, and not on
lapse and shift vector like {\it system IIa}.  We call this
system {\it system IIb}.

\setcounter{section}{5}
\section{Discussion} \label{sec:disc}

We have constructed several hyperbolic systems
based on the Ashtekar formulation of general relativity, 
together with discussions of the required gauge conditions and 
reality conditions.  We summarize their features in Table 
\ref{thetable}.

\begin{table}[h]
\begin{tabular}{c||c|c|c||c|c|c|c}
system & Eqs of  & reality  & gauge conditions  &
first   & all real  & diagonal- & sym.
\\
       & motion  &  condition & required &
order  & eigenvals & izable & matrix
\\
\hline \hline
{\it Ia} & original & metric & - & yes & yes & no & no
\\ \hline
{\it Ib} & original  & triad &
$\CA^a_0=\CA^a_iN^i$, $\ptl_i N=0$
  & yes & yes & no & no
\\ \hline
{\it IIa} & original & metric &
$N^l \neq 0, \,  \pm N \sqrt{\gamma^{ll}}$  ($\gamma^{ll}\neq 0$)
 & yes & yes & yes & no
\\ \hline
{\it IIb} & modified
& metric & $\CA^a_0=\CA^a_iN^i$  & yes & yes & yes & no
\\ \hline
{\it IIIa} & modified & triad & $\CA^a_0=\CA^a_iN^i$, \, $\ptl_i N=0$
& yes & yes & yes & yes
\\ 
\end{tabular}

\caption{List of obtained hyperbolic systems. The system {\it I, II}
and {\it III} denote weakly hyperbolic, diagonalizable
hyperbolic and symmetric hyperbolic systems, respectively. }
\label{thetable}
\end{table}

The original dynamical equations in the Ashtekar formulation
are classified as a weakly hyperbolic system. 
If we further assume a set of gauge conditions or reality 
conditions or both, then the system can be either 
a diagonalizable or a symmetric 
hyperbolic system.
We think such a restriction process 
helps in  understanding the structure of this dynamical system, 
and also that of the original Einstein equations. 
{}From the point of view of numerical applications, 
weakly and diagonalizable hyperbolic
systems are still good candidates to describe the spacetime dynamics
since they have much more gauge freedom than the 
obtained symmetric hyperbolic system. 

The symmetric hyperbolic system we obtained, is constructed  by
modifying the right-hand-side of the dynamical equations using 
appropriate combinations of the 
constraint equations. This is a modification of 
somewhat popular technique used also  
by Iriondo, Leguizam\'on and Reula.
We exhibited the process of determining coefficients, 
showing how uniquely they are determined
(cf Appendix \ref{appA}). 
In result, 
this symmetric hyperbolic formulation requires a triad reality condition,
which we suspect that Iriondo {\it et al} implicitly
assumed in their system.
As we demonstrated in \S \ref{sec5}, in order to keep the system first 
order, and to be consistent with the secondary triad reality 
condition, the lapse function is strongly restricted in form; 
it must be constant. 
The shift vectors and triad lapse ${\cal A}^a_0$ should have the
 relation
(\ref{shift-triad-relation}). This can be interpreted as the shift
being free and the triad lapse determined. 
This gauge restriction sounds tight, but this
arises from our general assumption of (\ref{A0katei}).
ILR propose to use the internal rotation to reduce this reality 
constraint, however this proposal does not work in our notation
(see Appendices \ref{appB} and \ref{appC}).

There might be a possibility to improve the situation
by renormalizing the shift and triad lapse terms into
the left-hand-side of the 
equations of motion like the case of
general relativity \cite{CBY}.
Or this might be
because our system is constituted by Ashtekar's original
variables.
We are now trying to relax this gauge restriction and/or
to simplify the characteristic speeds
by other gauge choices and also
by introducing new dynamical variables.
This effort will be reported elsewhere.


\vspace{0.4cm}
We thank John Baker for his useful comments on the initial draft.
We thank Abhay Ashtekar for his comments on our draft. 
We also thank Matt Visser for careful reading the manuscript. 
A part of this work was done when HS was at Dept. of Physics, 
Washington University, St. Louis, Missouri. 
HS was partially supported by NSF PHYS 96-00049, 96-00507,
and NASA NCCS 5-153 when he was at WashU.
HS was supported by the Japan Society for the Promotion of Science.

\appendix
\section{Detail processes of deriving the symmetric hyperbolic 
system IIIa}
\label{appA}

In this Appendix, we show several detail calculations 
for obtaining the symmetric hyperbolic system  {\it IIIa}.

\subsection{Determining $Q$ and $R$}\label{appA1}
We show here that 
the choice $Q$ and $R$ of (\ref{paraQR})
satisfies (\ref{d-d}). That is, the 
final $D$ in (\ref{fm-D}) satisfies  (\ref{cond2}),
and that this choice $Q$ and $R$ is unique.

First we show that
$D$ in (\ref{fm-D}) satisfies Hermiticity, (\ref{cond2}).
{}From the direct calculation, we get 
\beas
D^{labij} 
-
\bar{D}^{lbaji}
&=&
i \, \ut N (
 \ep^{abc} \dtri^j_c \gamma^{li}
-\ep^{abc} \dtri^l_c \gamma^{ji}
-\den^{-2}\ep^{bcd} \dtri^{ia}\dtri^j_c\dtri^l_d
+\den^{-2}\ep^{acd} \dtri^{jb}\dtri^i_c\dtri^l_d
-\den^{-2}\ep^{acd} \dtri^{lb}\dtri^i_c\dtri^j_d )
\\&&
+i \, \ut N (
 \ep^{bac} \dtri^i_c \gamma^{lj}
-\ep^{bac} \dtri^l_c \gamma^{ij}
-\den^{-2}\ep^{acd} \dtri^{jb}\dtri^i_c\dtri^l_d
+\den^{-2}\ep^{bcd} \dtri^{ia}\dtri^j_c\dtri^l_d
-\den^{-2}\ep^{bcd} \dtri^{la}\dtri^j_c\dtri^i_d )
\\&=&
i \, \ut N (
 \ep^{abc} \dtri^j_c \gamma^{li}
-\ep^{abc} \dtri^i_c \gamma^{lj}
-\den^{-2}\ep^{acd} \dtri^{lb}\dtri^i_c\dtri^j_d 
-\den^{-2}\ep^{bcd} \dtri^{la}\dtri^j_c\dtri^i_d 
) =: i \, \ut N  \dagger^{labij}. 
\enas
(This $\dagger^{labij}$ definition is used only within 
this subsection \ref{appA1}.)
Hermiticity, 
$\dagger^{labij}=0$, can be shown from the fact 
\beas
2\dagger^{l(ab)ij}
&=&
-\den^{-2}\ep^{acd} \dtri^{lb}\dtri^i_c\dtri^j_d 
-\den^{-2}\ep^{acd} \dtri^{lb}\dtri^j_c\dtri^i_d 
-\den^{-2}\ep^{bcd} \dtri^{la}\dtri^j_c\dtri^i_d 
-\den^{-2}\ep^{bcd} \dtri^{la}\dtri^i_c\dtri^j_d 
=0,
\enas
and its anti-symmetric part $\dagger^{l[ab]ij}=0$, which is
derived from 
\beas
\ep_{abe}\dagger^{labij}
&=&
 \ep_{abe}\ep^{abc} \dtri^j_c \gamma^{li}
-\ep_{abe}\ep^{abc} \dtri^i_c \gamma^{lj}
-\den^{-2}\ep_{abe}\ep^{acd} \dtri^{lb}\dtri^i_c\dtri^j_d 
-\den^{-2}\ep_{bea}\ep^{bcd} \dtri^{la}\dtri^j_c\dtri^i_d 
\\&=&
 2\dtri^j_e \gamma^{li}
-2\dtri^i_e \gamma^{lj}
-\den^{-2}\dtri^{lb}\dtri^i_b\dtri^j_e 
+\den^{-2}\dtri^{lb}\dtri^i_e\dtri^j_b 
-\den^{-2}\dtri^{la}\dtri^j_e\dtri^i_a 
+\den^{-2}\dtri^{la}\dtri^j_a\dtri^i_e 
\\&=&
 2\dtri^j_e \gamma^{li}
-2\dtri^i_e \gamma^{lj}
-\dtri^j_e \gamma^{il}
+\dtri^i_e \gamma^{lj}
-\dtri^j_e \gamma^{li}
+\dtri^i_e \gamma^{lj}
=0.
\enas

Next we show that
the choice $Q$ and $R$ of (\ref{paraQR})
is unique in order to satisfy (\ref{d-d}).
Suppose we have two pairs of ($Q, R$), say
($Q_1, R_1$) and ($Q_2, R_2$), as solutions of (\ref{d-d}).
Then the pair ($Q_1-Q_2, R_1-R_2$) should satisfy
a truncated part of (\ref{d-d}),
\begin{eqnarray}
\ddagger^{labij}
:=
-i \, Q^{ai} \ep^{bcd} \dtri^j_c \dtri^l_d
-i \, \bar{Q}^{bj} \ep^{acd} \dtri^i_c \dtri^l_d
-R^{ila} \dtri^{jb}
+R^{ija} \dtri^{lb}
+\bar{R}^{jlb} \dtri^{ia}
-\bar{R}^{jib} \dtri^{la}=0.
\label{katei}
\end{eqnarray}
Now we show that the equation $\ddagger^{labij}=0$ has only the
trivial solution $Q=R=0$.
By preparing
\bear
\ddagger^{labij}\gamma_{li}&=&
-iQ^{ai} \ep^{bcd} \dtri^j_c \dtri_{di}
-R^{ila}\gamma_{li} \dtri^{jb}
+R^{ija} \dtri^b_i,
\label{1}
\\
\ddagger^{labij}\gamma_{li}\tilde{E}_{jb}
&=&
-3\den^2 R^{ila}\gamma_{li}
+\den^2 R^{ija} \gamma_{ij}
=
-2\den^2 R^{ija}\gamma_{ij},
\nonumber
\enar
we get $
R^{ija}\gamma_{ij}=0
$.  By substituting this into (\ref{1}), we can express $R$ by $Q$ as
\be
R^{ija}
=
i \, \den^{-2} Q^{ak} \ep^{bcd} \dtri^j_c \tilde{E}_{kd}\dtri^i_b
=
i \, Q^{ak} \tilde{\ep}^{ij}{}_{k}.
\label{RQ}
\en

Therefore (\ref{katei}) becomes
$$
\ddagger^{labij}
=
-i \, Q^{ai} \ep^{bcd} \dtri^j_c \dtri^l_d
-i \, \bar{Q}^{bj} \ep^{acd} \dtri^i_c \dtri^l_d
-i \, Q^{ak} \tilde{\ep}^{il}{}_{k}  \, \dtri^{jb}
+i \, Q^{ak} \tilde{\ep}^{ij}{}_{k}  \, \dtri^{lb}
-i \, \bar{Q}^{bk} \tilde{\ep}^{jl}{}_{k}  \, \dtri^{ia}
+i \, \bar{Q}^{bk} \tilde{\ep}^{ji}{}_{k}  \, \dtri^{la}.
$$
{}From this equation, we get the following contracted relations:
\bear
\den^{-2}\ddagger^{labij}
&=&
-i \, Q^{ai} \ep^{bjl}
-i \, \bar{Q}^{bj} \ep^{ail}
-i \, Q^{ak} \ep^{il}{}_{k}  \, E^{jb}
+i \, Q^{ak} \ep^{ij}{}_{k}  \, E^{lb}
-i \, \bar{Q}^{bk} \ep^{jl}{}_{k}  \, E^{ia}
+i \, \bar{Q}^{bk} \ep^{ji}{}_{k}  \, E^{la}, 
\nonumber
\\
\den^{-2}\ddagger^{labij}E_{ia}
&=&
-i \, Q^a{}_a \ep^{bjl}
+2i \, Q^{ak} \ep_a{}^{[j}{}_{k}  \, E^{l]b}
-2i \, \bar{Q}^{bk} \ep^{jl}{}_{k},
\nonumber\\
\den^{-2}\ddagger^{labij}E_{ia}\ep_{ljc}
&=&
2i \, Q^a{}_a \delta^b_c
+2i \, Q^{bc}
-2i \, Q^{cb}
+4i \, \bar{Q}^{bc},
\label{ljc}
\\
\den^{-2}\ddagger^{labij}E_{ia}\ep_{ljc}\delta^c_b
&=&
6i \, Q^a{}_a
+4i \, \bar{Q}^a{}_a
=
10 i \, \re(Q^a{}_a)
-2 \im(Q^a{}_a).
\nonumber
\enar
where $Q^{ab}:=Q^{ai}E^b_i$
and $\ep^{bjl}:=\ep^{ijl}E^b_i$.
{}From the last one, we get $Q^a{}_a=0$. 
By substituting this into
(\ref{ljc}), we get
\be
\den^{-2}\ddagger^{labij} \, E_{ia} \ep_{lj}{}^c
=
2i \, Q^{bc}
-2i \, Q^{cb}
+4i \, \bar{Q}^{bc}
=
4i \, Q^{[bc]}
+4i \, \bar{Q}^{bc}
\label{kore}
\en
The symmetric part of (\ref{kore})
indicates
$Q^{(bc)}=0$,
and  
\beas
\den^{-2}\ddagger^{labij}E_{ja}E_{lb}
&=&
2i \, Q^{bc} \ep_{bc}{}^i 
-3i \, \bar{Q}^{bc} \ep_{bc}{}^i
=
-\re(Q^{bc}\ep_{bc}{}^i)
+5i \, \im(Q^{bc}\ep_{bc}{}^i)
\enas
gives us $Q^{bc}\ep_{bc}{}^i=Q^{[bc]}=0$.
Therefore
$Q^{bc}=Q^{ai}=0$ is determined uniquely.
{}From (\ref{RQ}), 
we also get $R^{ija}=0$.

\subsection{Modifications to off-diagonal blocks}\label{appA2}
On the starting point of the modifications to the equations of
motions (\ref{eqE2}) and  (\ref{eqA2}), we assumed that off-diagonal
terms keep vanishing.  In this subsection, we
show that the modifications to
the off-diagonal blocks $B$ and $C$ in the matrix notation of
(\ref{matrixform}), i.e.
modifying (\ref{eqE}) using ${\cal C}_{H}$ and  ${\cal C}_{M}$ and
modify (\ref{eqA})  using ${\cal C}_{G}$,
does not affect the final conclusion at all.

Suppose we have a symmetric hyperbolic system (\ref{fm-A})-(\ref{fm-D}),
and suppose we additionally modify the equations of motion (\ref{eqE})
and (\ref{eqA}) as
\bear
\mbox{modifying term for }\partial_t \tilde{E}^i_a 
&=&
G^i_a \CC_H +H^{ij}_a \CC_{Mj} \nonumber \\
&\cong&
G^i_a
(-i\ep_b{}^{cd}  \, \dtri^j_c \dtri^l_d)  (\ptl_l\CA^b_j)
+
H^{ik}_a
(-\delta^l_k \dtri^j_b
+\delta^j_k \dtri^l_b
)(\ptl_l\CA^b_j)
\nonumber
\\&=&
 (-iG^i_a\ep_b{}^{cd}  \, \dtri^j_c \dtri^l_d
-H^{il}_a  \dtri^j_b
+H^{ij}_a  \dtri^l_b  ) (\ptl_l\CA^b_j),
\\
\mbox{modifying term for }\ptl_t \CA^a_i&=&
I^{ab}{}_i  \, \CC_{Gb}
\cong
(I^{ab}{}_i  \, \delta^l_j)
(\ptl_l\dtri^j_b),
\enar
where $G^i_a$, $H^{ij}_a$ and $I^{ab}{}_i$ are parameters to be determined.
In the matrix notation, these can be written as
\bear
B^l{}_{ab}{}^{ij} &=&
-iG^i_a\ep_b{}^{cd}  \, \dtri^j_c \dtri^l_d
-H^{il}_a  \dtri^j_b
+H^{ij}_a  \dtri^l_b,
\\
C^{lab}{}_{ij}&=&
I^{ab}{}_i  \, \delta^l_j.
\enar
The Hermitian condition (\ref{cond3}) becomes
\bear
0&=&
-iG^i_a\ep_b{}^{cd}  \, \dtri^j_c \dtri^l_d
-H^{il}_a  \dtri^j_b
+H^{ij}_a  \dtri^l_b
-\bar{I}_{ba}{}^j  \, \gamma^{li} =: \dagger^l{}_{ab}{}^{ij} \label{daggerA2}
\enar
(We use this $\dagger^l{}_{ab}{}^{ij} $ definition only inside of this
subsection \ref{appA2}.)

If there exists a non-trivial combination of
$G^i_a$, $H^{ij}_a$ and $I^{ab}{}_i$
which satisfy this relation, then it will constitute
alternative symmetric hyperbolic system.
However, we see only the trivial solution is allowed 
for (\ref{daggerA2})
as follows.
{}From the relations of
$\dagger^{kabij}\gamma_{ij}+\dagger^{labik}\gamma_{li}
=-4\bar{I}^{bak}$,
we obtain $I^{ab}{}_i=0$.  With this $I^{ab}{}_i=0$, we obtain 
$\dagger^l{}_{ab}{}^{ij} \, \dtri^b_j=-2\den^2 H^{il}_a$,
which determine $H^{ij}_a=0$.
Similarly, from $I^{ab}{}_i=0$ and $H^{ij}_a=0$,
we get
$\dagger^{l}{}_a{}^{bij}  \, \ut\ep_{jlk} \dtri^k_b=-6i\den^2 G^i_a$,
which determine $G^i_a=0$.

\section{Internal rotation and Ashtekar equations}
\label{appB}
In this Appendix, 
we consider the effect of a $SO(3)$ rotation on the triad,
which corresponds to a $SU(2)$ rotation on the soldering form.
The equations that we derive here 
will be applied in the discussion in  Appendix \ref{appC}.

\subsection{Primary and secondary conditions of internal rotation}
The $SO(3)$ internal transformation only affects inner space, and
not the space-time quantities.  Let us write $U$ for such a rotation. 
$U$ should satisfy the condition
\be
U^a{}_c  \, U^{bc} = \delta^{ab}.
\label{unitary}
\en
This comes from the transformation of $\delta^{ab}$ to 
$\delta^{\ast ab}:=U^a{}_c \, U^b{}_d \, \delta^{cd}$, which should satisfy
$\delta^{\ast ab}=\delta^{ab}$.
The determinant $\det U$ must be $\pm 1$, and 
we choose $\det U=1$ for later convenience.
The transformation $\delta^a{}_b\to \delta^\ast{}^a{}_b$
is naturally defined by
$\delta^\ast{}^a{}_b:=U^a{}_c  \, U_b{}^d \, \delta^c{}_d$.
{}From (\ref{unitary}), we get the fundamental relations: 
$\delta^\ast{}^a{}_b=\delta^a{}_b$, 
$\delta^\ast{}_{ab}=\delta_{ab}$, 
and $\ep^\ast{}^{abc}=\ep^{abc}$. 

Now we define the transformation of the triad  $E^i_a$ 
and of the inverse triad $E^a_i$ as
\bear
E^\ast {}^i_a &:=& U_a{}^b  \, E^i_b. \label{triadH} \\
E^\ast {}^a_i &:=& U^a{}_b  \, E^b_i.
\enar
The 3-metric, $\gamma^{ij}$, 
is preserved under this transformation, since 
$\gamma^{ij}=E^i_a E^{ja}=E^\ast {}^i_a E^\ast {}^{ja}$. 
We note that this secondary condition,  
$\ptl_t \gamma^{ij}=\ptl_t (E^i_a E^{ja})=
\ptl_t (E^\ast {}^i_a E^\ast {}^{ja})$, will not give us
further conditions. This is equivalent with the time
derivative of (\ref{unitary}).

\subsection{Internal rotation of Ashtekar variables}
Using $\det U=1$, 
the transformation of the densitized triad becomes 
\bear
\dtri^\ast {}^i_a =
U_a{}^b  \, \dtri^i_b,
\label{tansdtri}
\enar
and straightforward calculation shows 
\bear
\CA^\ast{}^a_i
=
U^a{}_b  \, \CA^b_i
-\itwo \ep^{ab}{}_{c} \, U_b{}^{d} \, (\ptl_i U^c{}_d), 
\label{tansCA}
\enar
where we also note that 
$
\omega^\ast {}^{0a}_i = U^a{}_b  \, \omega^{0b}_i,
$ and $
\omega^\ast{}^{bc}_i =
 U^a{}_e(\ep^e{}_{bc}  \, \omega^{bc}_i)
 - \ep^a{}_{bc}(\ptl_i U^{bd})U^c{}_d.
$
We remark that the second term in (\ref{tansCA}) arises because 
$\CA^a_i$ includes the spatial derivative of the triad.
The relations of 
triad lapse and curvature 2-form become
\bear
\CA^\ast {}^a_0 &=& 
U^a{}_b  \, \CA^b_0
-\itwo \ep^{ab}{}_c  \, U_b{}^d(\ptl_t U^c{}_d)
, \\
F^\ast {}^a_{ij} &=& U^a{}_b   \, F^b_{ij}, 
\enar
and constraints (\ref{c-ham})-(\ref{c-g}) are transformed into 
\bear
\CC^\ast {}_H &=&  \CC_H, \\
\CC^\ast {}_{Mi} &=&  \CC_{Mi}, \\
\CC^\ast {}_{Ga} &=&  U_a{}^b  \, \CC_{Gb}.
\enar
The Hilbert action (\ref{action}) will be preserved ($S^\ast=S$)
under $U$, which is demonstrated by the ``cancellation relation"
\be
(\ptl_t \CA^\ast{}^a_i)\dtri^\ast{}^i_a
+\CA^\ast{}^a_0  \, \CC^\ast{}_{Ga}
=
(\ptl_t \CA^a_i)\dtri^i_a
+\CA^a_0 \, \CC_{Ga}.
\en
Therefore the equations of motion for 
$\dtri^\ast{}^i_a$ and $\CA^\ast{}^a_i$
are equivalent with the
original ones,  (\ref{eqE}) and (\ref{eqA}), putting a  
$\ast$ on all terms.

The secondary metric reality condition (\ref{w-reality2-final}), 
$W^{ij}:=\re (\ep^{abc}\dtri^\ast{}^k_a 
\dtri^\ast{}^{(i}_b \CD^\ast{}_k \dtri^\ast{}^{j)}_c)$, 
retains its form,  
\beas
W^\ast{}^{ij}=W^{ij},
\enas
while the 
secondary triad reality condition
(\ref{s-reality2-final2}), 
$Y^a:=-\re(\CA^a_0)+\ptl_i(N) E^{ia}+N^{i}\re(\CA^a_i)$, 
is transformed as
\bear
Y^\ast{}^a&=&
\re(U^a{}_b) Y^b
-i\ptl_i(N) \re(U^a{}_b) \im(E^i_b)
+\im(U^a{}_b)
[\im(\CA^b_0)
-\ptl_i(N) \im(E^{ib})
-N^{i} \, \im(\CA^b_i)] \nonumber 
\\&&
+\half  N^{i} \ep^a{}_{bc}  \, \im(U^{bd}) (\ptl_i  \, \im(U^c{}_d)).
\label{YScond_viaU}
\enar
This equation has many unexpected terms, even if we assume the 
triad reality, $\im(E^i_a)=0$,  before the transformation. 

To summarize, under triad transformations, 
$\CA^a_i$, $\CA^a_0$,  and  $Y^a$ are not 
transformed covariantly, while the  
other variables are
transformed covariantly.

\subsection{Make triad real using internal rotation}\label{appB3}
Suppose all the variables satisfy the metric reality conditions, that
is, $\dtri^i_a$ satisfies $\im(\dtri^i_a\dtri^{ja})=0$.
Can we obtain the triad which satisfies the triad reality 
condition, $\im(\dtri^\ast{}^i_a)=0$, by an internal rotation? 

The answer is affirmative. However, 
such a rotation $U$ must satisfy
\bear
0 =
\im(\dtri^\ast{}^i_a) =
\im(U_a{}^b \, \dtri^i_b) =
 \re(U_a{}^b) \, \im(\dtri^i_b)
+\im(U_a{}^b) \, \re(\dtri^i_b),
\enar
and its secondary condition
\be
0 =
\im(\ptl_t \dtri^\ast{}^i_a) =
\im[(\ptl_t U_a{}^b)\dtri^i_b
+U_a{}^b \, (\ptl_t \dtri^i_b)].
\label{YScond_viaU2}
\en

The application of this technique will be discussed in 
\S \ref{appC2}.  Before ending this section, we remark
two points. 
First, 
$\CA^a_i$ is not transformed covariantly by this rotation $U$. 
Second, 
when we consider the evolution of $\dtri^\ast{}^i_a$,
the evolution should be consistent with the 
secondary triad reality condition (\ref{s-reality2-final}).

\section{Consideration of ILR's treatment of reality conditions}
\label{appC}
The symmetric hyperbolic system (system {\it IIIa}) that we
obtained in \S \ref{sec5} is strictly restricted by the
triad reality condition. 
ILR (in their second paper \cite{ILRsecond}) propose to use an 
internal rotation to de-constrain this situation. 
Here we comment on this possibility. 

\subsection{Difference of definition of symmetric 
hyperbolic system}
First of all, we should point out again that there is a fundamental 
difference in  the definition used to characterize the system 
as {\it symmetric}. 
As we discussed in \S \ref{sec:def}, we define symmetry
 using the fact that
the characteristic matrix is Hermitian, while ILR
\cite{Iriondo,ILRsecond} define it when 
the principal symbol of the system $iB^l{}_j{}^a k_a$ 
($i\cha^{l\beta}{}_\alpha k_l$ in our notation)
is anti-Hermitian.

We suspect that these two definitions are equivalent 
when the vector $k_a$ ($k_l$ in our notation) 
is arbitrary real.  
Actually, ILR have advanced a suggestion 
that our definition and 
their `modern' version are equivalent.
The judgement which is conventional or not, however, 
we would like to leave to the reader. 
Concerning our definition of symmetric hyperbolicity, 
we think that the readers can quite easily compare our system with 
other proposed symmetric hyperbolic systems in general 
relativity:  
all eigenvalues (in the system we presented) are all real-valued, 
while ILR's are all pure imaginary.
(Even if the distinction of real and pure imaginary is ignored,
the eigenvalues calculated by us (\ref{symhypIIIa_eigen})
and by ILR are different.)

We note that, in addition, this fundamental difference will 
lead to different conclusions regarding the treatment of the 
reality condition  (see the proceeding discussion).


\subsection{Can we obtain a symmetric hyperbolic system by internal
rotation?}
\label{appC2}

What ILR proposed is the following: Suppose the system satisfies
the reality condition on the metric, but not on the triad. 
By using the freedom of making an internal rotation, we can transform 
the soldering form to satisfy the triad reality condition, in such 
a way it forms symmetric hyperbolic system. 
(In their terminologies, 
they seek a ``rotated" scalar product that is to 
find a more general symmetrizer.)
Therefore we can remove
the additional constraints of the triad reality. 

This procedure, however, includes changing inner product 
of dynamical variables, which might cause the topology 
of well-posedness of the initial value formulation to change.
Here, we examine whether such a re-definition of the inner product is
acceptable in our definition of symmetric hyperbolicity. 


Suppose we have a system which 
satisfies the constraints, and the metric reality condition, 
but not the triad reality conditions. As we commented in 
\S \ref{sec:ash}, 
metric reality will be preserved automatically by the dynamical 
equations (\ref{matrixform}) and (\ref{fm-A})-(\ref{fm-D}).
Now we apply a $SO(3)$ rotation
$E^i_a \to E^\ast {}^i_a:= U_a{}^b  \, E^i_b$ to the system. 
We summarized the transformations of Ashtekar's variables and 
equations by $U$ in Appendix \ref{appB}. In the new variables
$(\dtri^\ast{}^i_a,\CA^\ast{}^a_i)$,  transformed via $U$, 
the equations of  motions are written covariantly. 

As discussed in \S \ref{appB3}, it is possible to construct the
real triad by using $U$. However, we always should verify the triad
reality condition, both its primary condition (\ref{s-reality1}),
and its 
secondary condition (\ref{s-reality2}).
The latter is expressed as 
(\ref{YScond_viaU}) or (\ref{YScond_viaU2}). 
If we interpret this secondary condition as a restriction
on the gauge variables, lapse $N$, shift
$N^i$, and triad lapse $\CA^a_0$, then 
we only need to solve the primary condition in order to obtain 
triad reality on 3-hypersurface. 
This is indeed solvable. For example, ILR explain a way to get a real 
triad using orthonormality of the basis 
in their Appendix A in \cite{ILRsecond}. 

Next, let us see whether a symmetric hyperbolic system is obtained
by the new pair of variables $(\dtri^\ast{}^i_a,\CA^\ast{}^a_i)$.
We define the equations of motion similarly as 
\be
\partial_t \left[ \begin{array}{l}
\dtri^\ast{}^i_a \\
\CA^\ast{}^a_i
\end{array} \right] =
\left[ \begin{array}{cc}
A^\ast{}^l{}_a{}^{bi}{}_j & B^\ast{}^l{}_{ab}{}^{ij} \\
C^\ast{}^{lab}{}_{ij} & D^\ast{}^{la}{}_{bi}{}^j
\end{array} \right]
\partial_l
\left[ \begin{array}{l}
\dtri^\ast{}^j_b \\
\CA^\ast{}^b_j
\end{array} \right]
+ \mbox{terms~with~no~} \partial_l \dtri^\ast{}^j_b
\mbox{~nor~} \partial_l \CA^\ast{}^a_i.
\label{matrixform2}
\en
By applying the same modifications as those in  
\S \ref{sec5}, we get 
\bear
A^\ast{}^{labij}&=&
i\ep^{abc}  \ut N
\dtri^\ast{}^l_c \gamma^{ij}
+N^l\gamma^{ij} \delta^{ab},
\label{fm-A2}
\\
B^\ast{}^{labij}&=&C^\ast{}^{labij}=0,
\label{fm-BC2}
\\
D^\ast{}^{labij}&=&
i\ut N(\ep^{abc} \dtri^\ast{}^j_c \gamma^{li}
- \ep^{abc} \dtri^\ast{}^l_c \gamma^{ji}
 \nonumber \\ &~&
-e^{-2} \dtri^\ast{}^{ia}
 \ep^{bcd} \dtri^\ast{}^j_c \dtri^\ast{}^l_d
-e^{-2}\ep^{acd} \dtri^\ast{}^i_d
 \dtri^\ast{}^l_c  \dtri^\ast{}^{jb}
+e^{-2}
\ep^{acd} \dtri^\ast{}^i_d \dtri^\ast{}^j_c
  \dtri^\ast{}^{lb}
)
+N^l \delta^{ab} \gamma^{ij}.
\label{fm-D2}
\enar
These equations are related to (\ref{fm-A})-(\ref{fm-D}). We note that,
in the modification here, we added the terms
$
(N^i \delta^{ab}+
i\ut N \epsilon^{abc}\dtri^\ast{}^i_c)
\CC^\ast_{Gb}$
coming from the terms of the gauge constraint. This corresponds to the
relation $A^\ast{}^{labij}=
U^a{}_c \, U^b{}_d \, A^{lcdij}$.

Equations (\ref{fm-A2})-(\ref{fm-D2}) forms a Hermitian matrix 
in the principal part of (\ref{matrixform2}), but it contradicts
the consistent evolution with triad reality. 
That is, for example, the left-hand-side of dynamical
 equation $\ptl_t \dtri^\ast{}^i_a=\cdots$ [upper half of 
(\ref{matrixform2})]
is real-valued since we impose $\im(\dtri^\ast{}^i_a)=0$, while in the
right-hand-side includes complex value in the non-principal
part.  To explain this in another words, the system 
(\ref{matrixform2})-(\ref{fm-D2}) will not preserve the triad 
reality. Therefore we again need to control gauge variables through
the secondary triad reality condition, and this discussion again 
returns the same gauge restrictions with those in \S \ref{sec5}.

We also point out that the inner product of the fundamental variables
in our notation does not form Hermitian like in the case of ILR. 
The inner product before the rotation $U$ can be written 
\be
\langle 
(\dtri^i_a,\CA^a_i)| (\dtri^i_a,\CA^a_i) \rangle
:=
\delta^{ab}\gamma_{ij}\dtri^i_a\bar{\dtri}{}^j_b
+\delta_{ab}\gamma^{ij}\CA^a_i\bar{\CA}{}^b_j,
\en
which is common to ours and ILR's, while after the rotation the 
inner product becomes
\bear
\langle 
(\dtri^\ast{}^i_a,\CA^\ast{}^a_i)| 
(\dtri^\ast{}^i_a,\CA^\ast{}^a_i) \rangle 
&=&
U_c{}^a \, \bar{U}^{cb} \dtri^i_a \bar{\dtri}{}^j_b 
+
U_c{}^a \, \bar{U}^{cb} \gamma^{ij}\CA^a_i \bar{\CA}{}^b_j
\nonumber \\
&&
-\itwo\gamma^{ij} \left(
 \ep_{agf}\bar{U}{}^{gh}(\ptl_j \bar{U}{}^f{}_h)
 U^a{}_c \, \CA^c_i
+\ep^a{}_{ec} \, U^{ed}(\ptl_i U^c{}_d)
 \bar{U}{}_{af} \, \bar{\CA}{}^f_j
\right)
\nonumber \\
&&
-{1\over 4}U_e{}^d(\ptl_i U_{cd})
 \bar{U}{}^{eh}(\ptl_j \bar{U}{}^c{}_h)
+{1\over 4}U_e{}^d(\ptl_i U_{cd})
 \bar{U}{}^{ch}(\ptl_j \bar{U}{}^e{}_h),
\label{inner_product}
\enar
which is not Hermitian, and can not be used as
the inner product of the
original variable $(\dtri^i_a,\CA^a_i)$ as in the ILR's
proposal.

As the final remark, we would like to comment that both 
the variables to evolve by the equations,  and the variables used  
to confirm the Hermiticity of the system 
should be common
throughout all evolutions. Otherwise, we cannot apply the
energy inequality for the evolution of that system. 
{}From this point of view, we think it necessary to consider
the secondary triad 
reality condition throughout 
evolution of this system. 

To summarize, we tried to follow ILR's procedure to remove the
restriction of the triad reality condition in our system,
which casts on our definition of symmetric hyperbolicity, 
and which is based on the fixed inner product as of its Hermitian
form. 
We, however, see that ILR's procedure does not work in our system
since it requires the restriction of the secondary reality conditions
of the triad. 
Therefore we conclude that we cannot de-constrain 
restrictions any further.


[Additional Notes in this online version] 

Based on the results of this article,
 the authors developped a set of dynamical equations which 
forces the spacetime to evolve to the manifold that 
satisfies the constraint
equations or the reality conditions or both as the 
attractor against perturbative errors. This report is available as:
H. Shinkai and G. Yoneda, 
Phys. Rev. {\bf D60}, 101502 (1999). 

The authors also performed 
numerical comparisons of three levels of the hyperbolic forms 
obtained in this article.  The report is avilable as: 
H. Shinkai and G. Yoneda, 
gr-qc/0005003. 


\begin{thebibliography}{99}
\baselineskip .15in
\bibitem[\dagger]{Email-yone} Electronic address:
yoneda@mn.waseda.ac.jp
\bibitem[\ddagger]{Email-his}  Electronic address:
shinkai@gravity.phys.psu.edu

\bibitem{Reula98}
Recent review is given by
O. A. Reula,
Livng Rev. Relativ. {\bf 1998-3} at http://www.livingreviews.org/.

\bibitem{Heldbook}
Y. Choquet-Bruhat and J.W. York. Jr., in {\em General Relativity and
Gravitation }, vol. 1, ed. by A. Held, (Plenum, New York, 1980).


\bibitem{BonaMasso}
C. Bona, J. Mass\'{o}, 
Phys. Rev. Lett. {\bf 68}, 1097 (1992);
C. Bona, J. Mass\'{o}, E.. Seidel, J. Stela, 
Phys. Rev. Lett. {\bf 75}, 600 (1995);
Phys. Rev. {\bf D56}, 3405 (1997).

\bibitem{FischerMarsden72}
A. E. Fischer and J. E.  Marsden,
Commun. Math. Phys. {\bf 28}, 1 (1972).

\bibitem{ChoquetBruhat}
Y. Choquet-Bruhat and T. Ruggeri, 
Commun. Math. Phys. {\bf 89}, 269 (1983).

\bibitem{HF}
H. Friedrich, 
Proc. Roy. Soc. {\bf A375}, 169 (1981);
Proc. Roy. Soc. {\bf A378}, 401 (1981);
Comm. Math. Phys {\bf 100}, 525 (1985);
J. Diff. Geom. {\bf 34}, 275 (1991).


\bibitem{fried96}
H. Friedrich, Class. Quantum Grav. {\bf 13}, 1451 (1996).

\bibitem{VPE96} 
M.H.P.M. van Putten and D.M. Eardley,   
Phys. Rev. {\bf D53}, 3056 (1996).

\bibitem{CBY}
Y. Choquet-Bruhat and J.W. York, Jr.,  
C. R. Acad. Sci. Paris, {\bf t. 321}, 
S\'erie I, 1089 (1995);
A. Abrahams, A. Anderson, Y. Choquet-Bruhat and J.W. York, Jr.,
Phys. Rev. Lett. {\bf 75}, 3377 (1995);
C. R. Acad. Sci. Paris, {\bf t. 323}, 
S\'erie II b, 835 (1996);
Class. Quantum Grav. {\bf 14}, A9 (1997);
A. Anderson and  J.W. York, Jr. gr-qc/9901021.

\bibitem{FR94}
S. Frittelli and O.A. Reula, 
Comm. Math. Phys {\bf 166}, 221 (1994).

\bibitem{FR96}
S. Frittelli and O.A. Reula, 
Phys. Rev. Lett. {\bf 76}, 4667 (1996).

\bibitem{miguel}
M. Alcubierre, Phys. Rev. {\bf D55}, 5981 (1997).

\bibitem{SBCSThyper}
M.A. Scheel, T.W. Baumgarte, G.B. Cook, S.L. Shapiro, and
S.A. Teukolsky, Phys. Rev. {\bf D56} 6320 (1997);
{\it ibid.} {\bf D58}, 044020 (1998).

\bibitem{cactus1}
C. Bona, J. Mass\'o, E. Seidel and P. Walker, gr-qc/9804052.

\bibitem{Ashtekar}
A. Ashtekar, Phys. Rev. Lett. {\bf 57}, 2244 (1986);
             Phys. Rev.  {\bf D36}, 1587 (1987);
{\em Lectures on Non-Perturbative Canonical Gravity}
                             (World Scientific, Singapore, 1991).

\bibitem{ys-con}
G. Yoneda and  H. Shinkai,
Class. Quantum Grav. {\bf 13}, 783 (1996).

\bibitem{ysn-dege}
G. Yoneda, H. Shinkai and A. Nakamichi,
Phys. Rev. {\bf D56}, 2086 (1997).



\bibitem{Iriondo}
M.S. Iriondo, E.O. Leguizam\'on and  O.A. Reula,
Phys. Rev. Lett. {\bf 79}, 4732 (1997).


\bibitem{YShypPRL}
G. Yoneda and  H. Shinkai,
Phys. Rev. Lett. {\bf 82}, 263 (1999).

\bibitem{ILRsecond}
M.S. Iriondo, E.O. Leguizam\'on and O.A. Reula, 
Adv. Theor. Math. Phys.
{\bf 2}, (5), (1998)  [gr-qc/9804019].





\bibitem{weakhyperbolic}
See references in
E. Jannelli,
Comm. in Part. Diff. Eq., {\bf 14},  1617 (1989).

\bibitem{mizodiaghyp}
S. Mizohata, Mem. College Sci. Kyoto Univ. {\bf 32}, 181
(1959).


\bibitem{CH}
R. Courant and D. Hilbert, {\em Methods of Mathematical
Physics, Volume II}, (John Willey \& Sons, 1962).

\bibitem{stewart}
J.M. Stewart, Class. Quantum  Grav. {\bf 15}, 2865 (1998).
\bibitem{Gustaf}
B. Gustafsson, H-O. Kreiss and J. Oliger, {\em Time Dependent
Problems and Difference Methods}, 
(John Wiley \& Sons, New York, 1995)

\bibitem{Geroch}
R. Geroch, {\em Partial Differential Equations in Physics},
gr-qc/9602055. 

\bibitem{taniguchisymp} 
{\em 
Hyperbolic equations and related topics}, ed. by S.
Mizohata, (Academic Press, 1986)

\bibitem{AshtekarRomanoTate}
A. Ashtekar, J. D. Romano and R. S. Tate,
Phys. Rev.  {\bf D40}, 2572 (1989).

\bibitem{gr-qc/9902012}
M.S. Iriondo, E.O. Leguizam\'on and O.A. Reula, gr-qc/9902012.

\end{thebibliography}
\end{document}